\documentclass{iaus}
\usepackage{graphicx}

\title[Gas signatures of Herbig Ae/Be disks with Herschel SPIRE] %% give here short title %%
{Gas signatures of Herbig Ae/Be disks \\ probed with Herschel SPIRE spectroscopy}

\author[Matthijs H. D. van der Wiel et al.]   %% give here short author list %%
{
Matthijs H. D. van der Wiel$^1$,
David A. Naylor$^1$,
Giambattista Aresu$^2$,
G\"oran Olofsson$^3$
 }

\affiliation{
$^1$Institute for Space Imaging Science, Department of Physics and Astronomy, University of Lethbridge, Lethbridge, AB, Canada; email: {\tt matthijs.vanderwiel@uleth.ca} \\[\affilskip]
$^2$Kapteyn Astronomical Institute, University of Groningen, Groningen, The Netherlands \\[\affilskip]
$^3$Stockholm University, Stockholm, Sweden
}

\pubyear{2013}
\volume{299}  %% insert here IAU Symposium No.
\pagerange{xxx--xxx}
\jname{Exploring the formation and evolution of planetary systems}
\editors{B. Matthews, \& J. Graham, eds.}
\begin{document}

\maketitle

\begin{abstract}
Herbig Ae/Be objects, like their lower mass counterparts T Tauri stars, are seen to form a stable circumstellar disk which is initially gas-rich and could ultimately form a planetary system. We present {\it Herschel} SPIRE 460--1540 GHz spectra of five targets out of a sample of 13 young disk sources, showing line detections mainly due to warm CO gas. 
\end{abstract}

\firstsection % if your document starts with a section,
\section{Observations, processing, and line detections}

\begin{figure}[htb]
% \vspace*{-2.0 cm}
\begin{center}
 \includegraphics[width=1.0\textwidth]{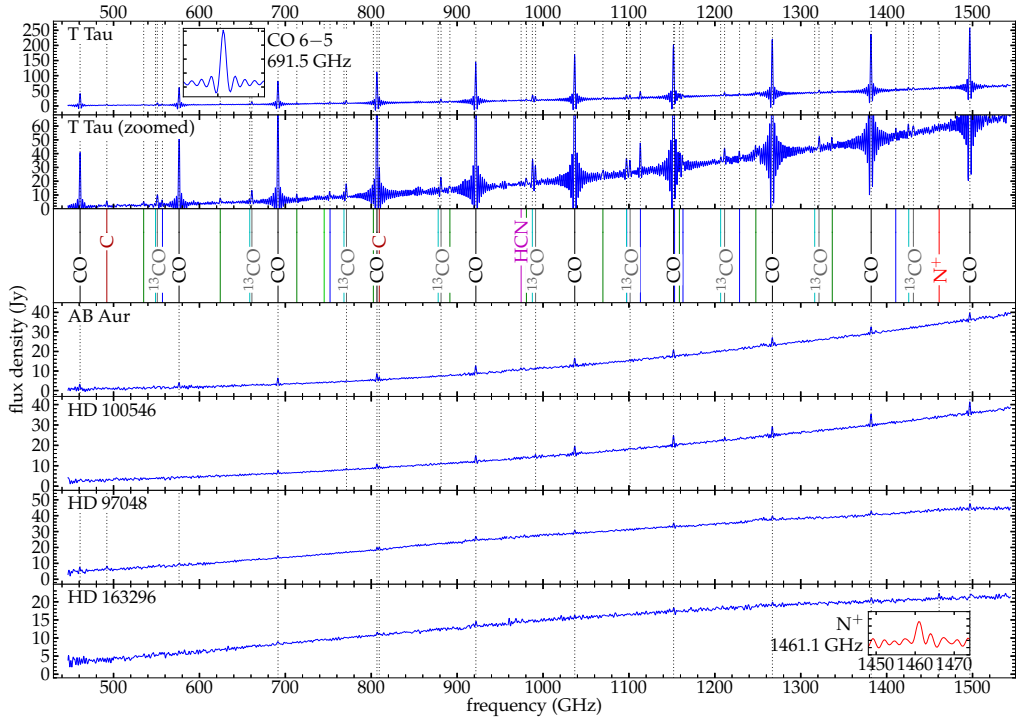} 
 \vspace*{-0.3 cm}
 \caption{{\it Herschel} SPIRE-FTS spectra of T\,Tau (with the second panel zoomed in on the flux density axis to highlight weaker lines), AB\,Aur, HD\,100546, HD\,97048 and HD\,163296. Detected gas lines are indicated by vertical dotted lines. The ringing of the sinc-shaped line profile typical of an FTS (see inset in top panel) is particularly noticeable near the $^{12}$CO lines in T\,Tau. }
   \label{fig:spectra}
\end{center}
\end{figure}

In a 16-hour guaranteed time project (P.I. G\"oran Olofsson) using the SPIRE Fourier Transform Spectrometer (FTS), spectra in the 460--1540\,GHz range were obtained of 13 protoplanetary disks around Herbig stars: AB\,Aur, HD\,100546, HD\,97048, HD\,163296, T\,Tau, HD\,142527, HD\,144432, RY\,Tau, HD\,104237, HD\,36112, HD\,169142, HD\,100453 and TW\,Hya. The data were processed in HIPE 9, 
%A 16-hour SPIRE guaranteed time project (P.I. G\"oran Olofsson) was devoted to obtain 460--1540 GHz spectra of 13 protoplanetary disks around Herbig stars: AB\,Aur, HD\,100546, HD\,97048, HD\,163296, T Tau, HD\,142527, HD\,144432, RY\,Tau, HD\,104237, HD\,36112, HD\,169142, HD\,100453 and TW\,Hya. Processing of the Fourier Transform Spectrometer (FTS) data is done with HIPE 9, 
followed by subtraction of a median background signal obtained from the off-center detectors. 
All spectra show a smoothly rising dust continuum, while only the first five of the above list show detectable line signal. 
The disk emission lines, presented in Fig.~\ref{fig:spectra}, are unresolved both spatially (the SPIRE FTS beam is 42--17$''$) and spectrally ($R$ $\sim$ 400--1300). 
Spectral lines are fitted using a dedicated Fourier Transform line fitter tool (available online at \texttt{www.uleth.ca/phy/naylor/}). Table~\ref{t:linedetections} lists lines detected in at least one target besides T Tau.

\section{Interpretation and analysis plans}

First, it is evident from the observations that the continuum and CO lines toward T\,Tau originate in an extended protostellar envelope rather than in a much smaller disk. This is likely also the case for the cold H$_2$O vapor.  
Second, the N$^+$ line toward HD\,163296 is $10^4$--$10^5$ times brighter than what is predicted by current (X-ray) irradiated disk models (e.g., Aresu \etal\ 2011), even when invoking X-ray luminosities much higher than appropriate for HD\,163296. Since N$^+$ is also detected in all off-center detectors up to 1$'$ away from the star, we hypothesize that a single (external) source is responsible for ionizing nitrogen both in the disk/jet of HD\,163296 and in the surrounding gaseous medium.
Finally, we plan to use the unique set of 4--3 to 13--12 lines of CO and $^{13}$CO ($E_\mathrm{up}/k$ $\sim$ 50--500\,K) to characterize warm gas in the Herbig protoplanetary disks. \\

\textit{Acknowledgements:} This work is supported by CSA and NSERC. SPIRE has been developed by a consortium of institutes led by Cardiff University (UK) and including contributions from Canada, China, France, Italy, Spain, Sweden, UK, and USA.

\begin{table}[!htb]
  \begin{center}
  \caption{Spectral lines identified in the SPIRE spectra of at least one target besides T\,Tau.}
  \label{t:linedetections}
 {\scriptsize
  \begin{tabular}{l r r @{\hspace{2em}} r r r r r}
  \hline 
  {\bf line } 	& {\bf $E_\mathrm{up}/k$}	& {\bf frequency}$^\mathrm{(2)}$	& \multicolumn{5}{c}{{\bf line flux}$^\mathrm{(3)}$ ($10^{-18}$ W\,m$^{-2}$) {\bf [uncertainty]}} \\
\cline{4-8} \\[-0.6em]
\hspace{0.5em} {\bf transition}$^{(1)}$	& {\bf (K)} & {\bf (GHz)}	& {\bf T\,Tau} & {\bf AB\,Aur} & {\bf HD\,100546} & {\bf HD\,97048} & {\bf HD\,163296} \vspace{-0.2em} \\
\hline
CO 4--3			& 55	 & 461.0 & 448 [5] & 35 [2] & - & 37 [3] & - \\
C $^3$P$_1$--$^3$P$_0$ & 24 & 492.2 & 35 [5] & - & - & 26 [3] & - \\
CO 5--4			& 83	 & 576.3	& 599 [5]	& 25 [2] & - & 13 [3] & - \\
CO 6--5			& 116 & 691.5 & 740 [5] & 38 [2] & 17 [2] & 17 [3] & 8 [2] \\
$^{13}$CO 7--6	& 148 & 771.2 & 100 [5] & 6 [2] & 7 [2] & - & - \\
CO 7--6			& 155 & 806.7 & 1141 [5] & 38 [2] & 22 [2] & 17 [3] & 10 [2] \\
C $^3$P$_2$--$^3$P$_1$ & 62 & 809.3 & 93 [5] & 14 [2] & 8 [2] & 17 [3] & 5 [2] \\
CO 8--7			& 199 & 921.8 & 1531 [5] & 48 [2] & 36 [2] & 31 [3] & 16 [2] \\
HCN 11--10		& 281 & 974.5 & 10 [4] & 11 [2] & - & - & - \\
$^{13}$CO 9--8	& 238 & 991.3 & 75 [4] & 5 [2]  & 14 [2] & - & - \\
CO 9--8			& 249 & 1036.9 & 1106 [4] & 47 [2] & 45 [2] & 25 [3] & 9 [2] \\
CO 10--9			& 304 & 1152.0 & 1202 [4] & 39 [2] & 53 [2] & 29 [3] & 12 [2] \\
$^{13}$CO 11--10	& 349 & 1211.3 & 50 [4] & - & 14 [2] & - & - \\
CO 11--10		& 365 & 1267.0 & 1225 [4] & 46 [2] & 51 [2] & 32 [3] & 12 [2] \\
CO 12--11		& 431 & 1382.0	 & 1224 [4] & 38 [2] & 59 [2] & 24 [3] & 10 [2] \\
N$^+$ $^3$P$_1$--$^3$P$_0$ & 70 & 1461.1 & - & - & - & - & 17 [2] \\
CO 13--12		& 503 & 1496.9 & 1262 [4] & 44 [2] & 62 [2] & 41 [3] & 11 [2] \\
\hline
  \end{tabular}
  }
 \end{center}
%\vspace{1mm}
 \scriptsize{
 {\it Notes:} 
 $^{(1)}$ In addition to the lines listed here, in T Tau we detect six H$_2$O lines and $\sim$25 others identified as $^{13}$CO, C$^{18}$O, HCO$^+$, HCN and N$_2$H$^+$. 
 $^{(2)}$ Rest frequency from JPL, \cite{pickett1998}, except N$^+$ from SLAIM. 
 $^{(3)}$ A `-' indicates a non-detection. Uncertainties include only the formal fitting error.  
}
\vspace{-0.5em}
\end{table}

\end{document}